# 2D correlations in the van der Waals ferromagnet $CrBr_3$ using high frequency electron spin resonance spectroscopy

C. L. Saiz[1], J. A. Delgado[1], J. van Tol[2], T. Tartaglia[3], F. Tafti[3], S. R. Singamaneni[1*]

**AFFILIATIONS**

[1]Department of Physics, The University of Texas at El Paso, El Paso, Texas 79968, USA

[2]National High Magnetic Field Laboratory, Florida State University, Tallahassee, Florida 32310, USA

[3]Department of Physics, Boston College, Chestnut Hill, Massachusetts 02468, USA

*Corresponding author email: srao@utep.edu

**ABSTRACT**

Broadening the knowledge and understanding on the magnetic correlations in van der Waals layered magnets is critical in realizing their potential next-generation applications. In this study, we employ high frequency (240 GHz) electron spin resonance (ESR) spectroscopy on plate-like $CrBr_3$ to gain insight into the magnetic interactions as a function of temperature (200 – 4 K) and the angle of rotation θ. We find that the temperature dependence of the ESR linewidth is well described by the Ginzberg-Landau critical model as well as Berezinskii-Kosterlitz-Thouless (BKT) transition model, indicative of the presence of two-dimensional (2D) correlations. This suggests that the three-dimensional ferromagnet $CrBr_3$, which has been described as an Ising or Heisenberg ferromagnet, could present 2D magnetic correlations and BKT-like behavior even in its bulk form; an observation that, to the best of our knowledge, has not been reported in the literature. Furthermore, our findings show that the resonance field follows a $(3cos^2\theta - 1)$-like angular dependence, while the linewidth follows a $(3cos^2\theta - 1)^2$-like angular dependence. This observed angular dependence of the resonance field and linewidth further confirm an unanticipated 2D magnetic behavior in $CrBr_3$. This behavior is likely due to the interaction of the external magnetic field applied during the ESR experiment that allows for the mediation of long-range vortex-like correlations between the spin clusters that may have formed due to magnetic phase separation. This study demonstrates the significance of employing spin sensitive techniques such as ESR to better understand the magnetic correlations in similar van der Waals magnets.

## I. INRODUCTION

Among various van der Waals (vdW) magnets, $CrX_3$ (where X = Cl, Br, I) is a family of compounds that has gained considerable attention due to their cleavable nature and persistent magnetic properties even at the atomic limit [1-5]. Of these, $CrBr_3$ is a soft out-of-plane ferromagnetic insulator with a Curie temperature ($T_C$) of about 33 K [5,6] and has been reported to show spontaneous magnetization down to the monolayer [7]. It is an interesting platform to study the magnetic interactions as atomically thin $CrBr_3$ has been shown to exhibit ferromagnetic interlayer coupling [8], while recent spin-polarized scanning tunneling microscopy and spectroscopy studies have confirmed this coupling to either be ferromagnetic (FM) or antiferromagnetic (AFM) in the bilayer depending on the stacking order [1]. Basic magnetic properties of this title compound have been reported earlier, with the magnetism arising from the $Cr^{3+}$ ion (S = 3/2, spin-only contribution) and the easy axis of magnetization along the *c*-axis [9,10].

Electron spin resonance (ESR) spectroscopy is an ideal tool to study the magnetic interactions and magnetic anisotropy in vdW magnetic materials [11-15]. For instance, through the use of high-frequency (*ν* = 240 GHz) ESR spectroscopy, Lee and co-authors have reported on the fundamental spin interactions that cause the magnetic anisotropy in $CrI_3$ [11]. While in another study, ESR spectroscopy was carried out at various frequencies in the presence of a high magnetic field in order to investigate the magnetic anisotropy in another vdW magnet, $Cr_2Ge_2Te_6$ [12]. In another recent work, Zeisner and co-authors again used high-field ESR spectroscopy over a range of frequencies to prove the existence of ferromagnetic short-range correlations above the magnetic phase transition and established the intrinsically two-dimensional (2D) nature of the magnetism in $CrCl_3$ [13]. Additionally, our group has employed light-induced ESR spectroscopy to study the photoexcited magnetic interactions in $CrI_3$ and $CrCl_3$ [15].

While numerous ESR works have been carried out on two compounds ($CrI_3$ and $CrCl_3$) of the $CrX_3$ family, to our knowledge, there has been only a few ESR works reported on $CrBr_3$ in the last 50 years [16,17]. However, the temperature dependence of the ESR linewidth, as well as the angular dependence of the ESR resonance field and linewidth have not been discussed in detail. The temperature and angular dependences of the ESR spectral parameters such as signal width and resonance field provide fingerprint signatures of magnetic phase transitions as well as information on microscopic magnetic interactions, which appears to be missing from the literature. Since the magnetic resonance properties of bulk $CrBr_3$ serve as a basis for the understanding of magnetic phenomena in reduced dimensions, e.g., in few- or monolayer samples of $CrBr_3$, a profound understanding of magnetic correlations in CrBr$_3$ is needed, especially considering that magnetic mono/few-layers of CrBr$_3$ have now become accessible. Therefore, it is important to expand the knowledge on this compound further. Here, through the use of high frequency ($v$ = 240 GHz) ESR spectroscopy, we show the temperature dependence of the ESR linewidth, reflective of spin dynamics, follows the Berezinskii-Kosterlitz-Thouless (BKT) transition model satisfactorily, while the Ginzberg-Landau critical model gives critical exponents indicative of two-dimensionality. In addition, it is found that the angular dependence of the linewidth and resonance field from CrBr$_3$ point toward 2D correlations. We explain this behavior as the coexistence of ferromagnetism and antiferromagnetism in the bulk CrBr$_3$ sample, similar to bulk CrI$_3$ [1,18,19].

## II. EXPERIMENTAL DETAILS

CrBr$_3$ single crystals of dimension 1 x 1 mm were grown by the CVT method described in a previous report [20,21]. High frequency ESR measurements were conducted at the National High Magnetic Field Laboratory (MagLab) located at Florida State University using the quasioptical spectrometer developed on-site. This system uses a superheterodyne spectrometer, employing a quasioptical submillimeter bridge that operates in reflection mode without cavity using a sweepable 17 T superconducting magnet [14]. In order to conduct ESR measurements as a function of angle (θ, degrees) at 200 and 4 K, sample tubes were loaded into the ESR cavity while attached to a sample rotator, allowing the sample to be set to various orientations within the cavity. The sample was cooled from room temperature down to 200 K before a magnetic field was applied in order to ensure that the vacuum grease used to attach the sample to the rotator is solidified. ESR measurements were performed at 240 GHz as a function of θ at both 200 K (paramagnetic phase) and 4 K (ferromagnetic phase) while the plate-like sample was rotated inside the cavity. Measurements were also conducted while the sample was held at a fixed θ of 70°, due to an apparent field minimum at that angle, while sweeping the temperature range from 200 – 4 K (as shown in Fig. 1).

## III. RESULTS AND DISCUSSION

Fig. 1 shows the 240 GHz ESR spectra collected at the fixed angle of θ = 70° over the temperature range of 200 – 4 K in order to gain insight into the behavior of the linewidth (full width at half maximum) and resonance field.

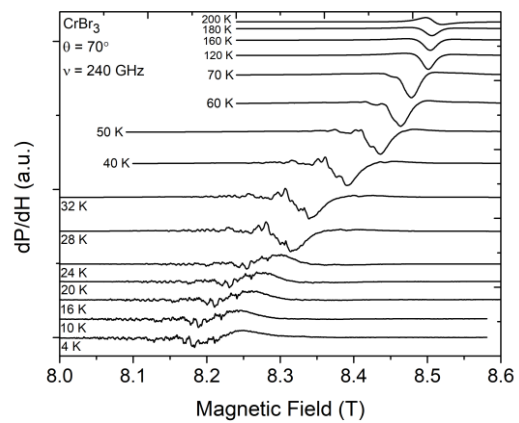

FIG. 1. ESR signals as a function of temperature at a single fixed angle.

It is observed that the single paramagnetic signal of Lorentzian line shape at 200 K with $g$ = 2.016 begins to split into multiple signals as the sample is cooled from 200 – 4 K. The temperature at which the single signal becomes multiple is in the vicinity of $T_C$ = 33 K. Below T = 32 K down to 4 K, multiple signals arise, and the paramagnetic signal broadens. Due to the high sensitivity of ESR spectroscopy compared to conventional magnetometers, this technique is able to pick up ferromagnetic correlations much above $T_C$ as reflected through the appearance of multiple signals even above 32 K, consistent with the previous reports [13]. Numerous ESR signals are typically observed in ferromagnetic compounds due to magnetocrystalline anisotropy, magnetic inhomogeneity and magnetic phase separation [22-27]. The $g$-value is seen to increase at a somewhat exponential rate as temperature is lowered from $g$ = 2.016 at T = 200 K to $g$ = 2.082 at T = 4 K beginning at T = 40 K (not shown). This temperature is within close proximity to $T_C$, suggesting that this sharp change in $g$-value can be attributed to the creation of internal magnetic correlations as the temperature approaches the magnetic phase transition from above.

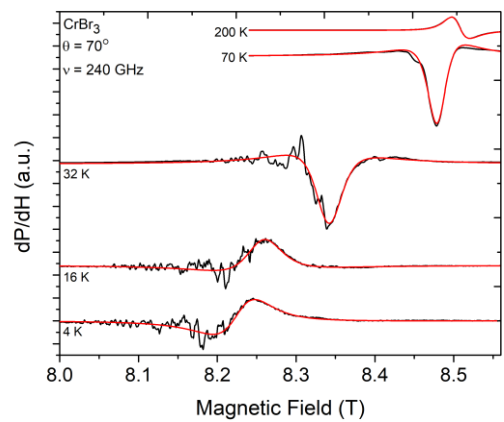

FIG. 2. Computer-generated fits employing a Lorentzian line shape over a range of temperatures represented by the smooth red line. The black line is the collected ESR spectra.

**A. Ginzberg-Landau Critical Model and BKT Transition**

The ESR spectral parameters were extracted using Lorentzian fits at each temperature in the paramagnetic phase (40 - 200 K), some of which are shown in Fig. 2. It should be noted that the narrow ESR signals modulated by a broader signal for the spectra measured at and below the magnetic phase transition were not fitted due to their apparent complexity. We first discuss the temperature dependence of the ESR linewidth as plotted in Fig. 3. Fig. 3(a) shows the variation of linewidth as a function of temperature represented by solid squares, while the superimposed red curve is the computer-generated fit resulted from the Ginzberg-Landau critical model. The critical model states,

$$\Delta H(T) = \frac{Q}{\left(\frac{T}{T_C}-1\right)^p} + mT + H_0 \quad — (1)$$

in which $Q$ is an arbitrary constant of proportionality, $T_C$ can refer to the Curie temperature, and $p$ is a critical exponent which is dependent on the spatial and spin degrees of freedom and can give a clear description of the dimensionality [28]. Fig. 3(b) shows the variation of linewidth as a function of temperature along with the fit generated using the BKT model, a phase transition that takes place at finite temperatures previously understood to be unique to the 2D XY model, a limiting case of the 2D Heisenberg model [5]. A transition of this type is not usually observed in condensed matter systems due to inherent interlayer coupling which inhibits the BKT transition due to long-range order [28], a property that 2D Heisenberg magnets lack at finite temperatures [3]. The BKT transition, which is typically found in 2D systems approximated by the XY model asserts,

$$\Delta H(T) = \Delta H_\infty \exp\left[\frac{3b}{\sqrt{\left(\frac{T}{T_{BKT}}-1\right)}}\right] + mT + H_0 \quad — (2)$$

In our experiment, $b$ has been set to $\pi/2$ for a square lattice, although it has been shown to remain valid at any value [29]. Fig. 3(c) shows the comparison in the fits generated using the two models. Fig. 3 illustrates that our obtained ESR data is described very nicely by Eq.'s (1) and (2) in the region well above the magnetic phase transition (see Table 1 for extracted fit parameters), with the extrapolated critical exponent $p$ = 0.9. This is very close to the reported value of $p$ = 0.5 – 0.7, which has been reported in most of the previously investigated layer-type antiferromagnets studied by Benner and Boucher [30]. A similar BKT-like transition has been observed in the $Cr^{3+}$-doped three-dimensional (3D) magnets such as $Bi_{0.5}Sr_{0.5}Mn_{0.9}Cr_{0.1}O_3$ by Ashoka and co-authors. Through the temperature dependence of the ESR linewidth, the authors employed the BKT model to satisfactorily describe their experimental findings. The authors explained this observation in terms of an effective 2D XY easy plane anisotropy induced by the magnetic field applied in the ESR experiment that allows for the mediation of long-range vortex-like correlations between spin clusters formed due to phase segregation [28]. Here, we will try to understand how bulk 3D $CrBr_3$ may exhibit magnetic phase separation, which has been known as a single phase ferromagnet. Quite interestingly, in a recent work reported by Li and co-authors [19], using Raman spectroscopy mapping, it has been revealed a novel mixed state of layered AFM and FM in 3D $CrI_3$ bulk crystals where the layered AFM survives in the surface layers, and the FM appears in deeper bulk layers. Similarly, $CrBr_3$ is also expected to show magnetic phase separation. With the applicability of BKT model on $CrBr_3$ demonstrated, we have shown that there is ample evidence of 2D correlations present within layered $CrBr_3$. This is argued to be caused by the presence of both FM and AFM correlations within a single bulk sample, as well as a consequence of the spectrometer's applied magnetic field combined with the inherent anisotropy in $CrBr_3$ [1,19,28,29,31].

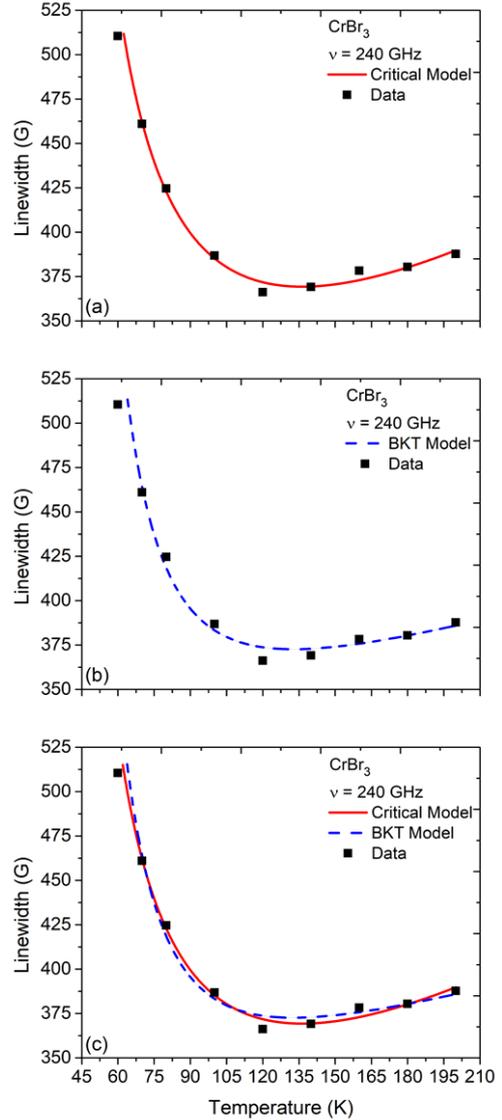

FIG. 3. Temperature dependence of the ESR linewidth at the fixed angle θ = 70°; fitted to the (a) Ginzberg-Landau

critical model and (b) Berezinskii-Kosterlitz-Thouless model (c) shows both the critical (solid red line) and BKT model (dashed blue line).

TABLE 1. Values of fit parameters and goodness of fit obtained from the Ginzberg-Landau Critical and BKT model on $CrBr_3$ single crystal.

| Critical model | | BKT model | |
|---|---|---|---|
| Exponent From Fit | | Exponent From Fit | |
| $Q$ (T) | 294.1 | $H_0$ (T) | 117.5 |
| $p$ | 0.9043 | $T_{BKT}$ (K) | 13.74 |
| $T_C, T_N$ (K) | 32 | $b$ | $\pi/2$ |
| $m$ (T/K) | 0.8759 | $H_\infty$ (T) | 27.64 |
| $H_0$ (T) | 148.9 | $m$ (T/K) | 0.8641 |
| $R^2$ | 0.9903 | $R^2$ | 0.9906 |

**B. Angular Dependence of the Resonance Field**

We now focus on the angular dependence of the ESR resonance field and linewidth. In Fig. 4(a), we have plotted the resonance field (solid squares, left y-axis) as well as the *g*-value (hallow inverted triangles, right y-axis) as a function of angle collected at 200 K (in the paramagnetic phase). The *g*-values were interpolated with a spline function. The angular dependence of the resonance field is fitted with the model,

$$H_{res}(\theta) = F(3\cos^2\theta - 1) + G \text{ — (3)}$$

The observed $(3\cos^2\theta - 1)$-like behavior of the resonance field is a characteristic behavior of 2D magnetic systems [32,33] and has earlier been observed in $K_2MnF$, $K_2CuF$, as well as $CrCl_3$, which has been experimentally shown to exhibit 2D Heisenberg behavior and is reported to have correlations suggestive of a weak XY model [19,31,32]. The curve connecting the resonance field values result from the fit using Eq. (3). An identical trend is observed for the angular dependence of resonance field and *g*-value when the measurement is conducted at T = 4 K (in the ferromagnetic phase), as plotted in Fig. 4(b). This marked angular dependence of the resonance field at both temperatures has been described as the noncubic distribution of the dipoles within the 2D lattice where the resulting net dipolar field shifts the resonance field according to Eq. (3) [32].

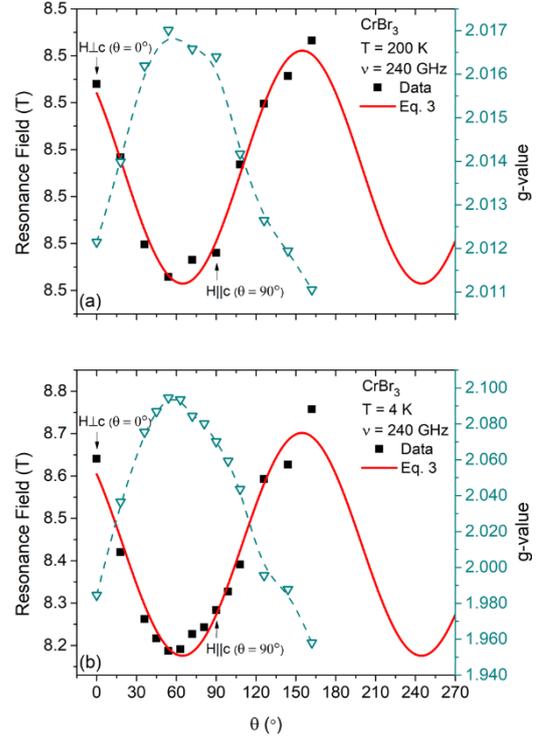

FIG. 4. Angular dependence of the resonance field where $v$ = 240 GHz at: (a) 200 K and (b) 4 K fitted with a modified form of Eq. (3) to take into account a phase shift: $H_0(\theta) = F[3\cos^2(\theta - \phi) - 1] + G$. Here, the cyan data points show the calculated *g*-values with a B-spline to guide the eye and illustrate the inverse relationship between resonance field and *g*-value.

**C. Angular Dependence of the Linewidth**

Equation (4) accurately and adequately describes the dependence of the linewidth on θ, both in ferromagnetic (4 K) as well as in the paramagnetic (200 K) phases as seen in Fig. 5,

$$\Delta H(\theta) = A(3\cos^2\theta - 1)^2 + B \text{ — (4)}$$

The nearly "W"-shaped angular dependence of the linewidth seen here is unique to low-dimensional magnets and has been previously reported in $MnPS_3$, a compound that has been shown to have 2D characteristics [34], as well as other 2D magnetic systems such as the 2D antiferromagnet $K_2MnF_4$ [30] and the out-of-plane 2D Heisenberg antiferromagnet $CrCl_3$ [32], which, as mentioned previously, is closely related to $CrBr_3$ due to their structural similarities. The unique shape of this curve with a maximum near lower angles, and a shallow minimum near θ = 55°, or the "magic" angle, has been observed in other systems and is a known feature of low-dimensional systems [30]. Similar to the model described by Eq. (3), this $(3\cos^2\theta - 1)^2$-like behavior is also characteristic of 2D magnetic systems and emerges from the dominant effects of the q ~ 0 modes in the long-time diffusional decay of the spin correlation function in such systems [32].

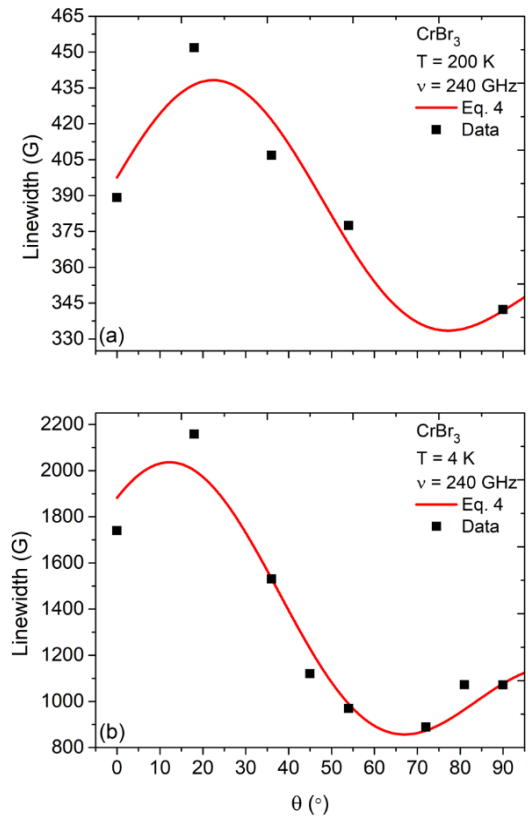

FIG. 5. Angular dependence of the linewidth at frequency $v$ = 240 GHz at: (a) 200 K and (b) 4 K fitted with a modified form of Eq. (4) to take into account a phase shift: $\Delta H(\theta) = A[3\cos^2(\theta - \phi) - 1]^2 + B$.

## IV. CONCLUSION

To conclude, we have conducted high frequency ESR spectroscopy on the single crystal chromium halide $CrBr_3$ as a function of temperature from T = 200 – 4 K and as a function of angle from θ = 0 – 162° to gain insight into the effects of temperature, as well as the angular dependence of the linewidth and resonance field. Using various models and fits, we determine that the 3D vdW ferromagnet $CrBr_3$ can exhibit 2D correlations by the BKT transition, likely due to the coexistence of ferromagnetism and antiferromagnetism in the bulk $CrBr_3$ sample. Similar to the magnetic correlations recently discovered in $CrI_3$, this uncommon behavior is thought to arise from the stacking structure and interlayer AFM correlations within the sample. It is also understood to be a consequence of the spectrometer's applied magnetic field combined with $CrBr_3$'s characteristic anisotropy. With these findings we have illustrated the importance of employing high sensitivity spectroscopy via ESR and suggest that $CrBr_3$ may require additional investigation to better understand its magnetic correlations that may extend to other layer-type vdW magnets.

## ACKNOWLEDGEMENTS

C. L. Saiz and S. R. Singamaneni acknowledge support from the UTEP start-up grant. T. Tartaglia and F. Tafti acknowledge NSF-DMR 1708929.